\documentclass[preprint,a4paper,superscriptaddress,floatfix]{revtex4-1}
\usepackage[english]{babel}
\usepackage[utf8]{inputenc}
\usepackage{amsmath}
\usepackage{hyperref}
\usepackage{graphicx,epstopdf,color}
\usepackage{xspace}

\def\linio{LiNiO$_2$\xspace}

\begin{document}

\title{Paraorbital ground state of trivalent Ni ion in LiNiO$_2$ from DFT+DMFT calculations}

\author{Dm.M.~Korotin}
\affiliation{M.N. Mikheev Institute of Metal Physics, S.Kovalevskoy St. 18, 620108 Yekaterinburg, Russia}
\email{dmitry@korotin.name}

\author{D.~Novoselov}
\affiliation{M.N. Mikheev Institute of Metal Physics, S.Kovalevskoy St. 18, 620108 Yekaterinburg, Russia}
\affiliation{Institute of Physics and Technology, Ural Federal University, Mira St. 19, 620002 Yekaterinburg, Russia}

\author{V.I.~Anisimov}
\affiliation{M.N. Mikheev Institute of Metal Physics, S.Kovalevskoy St. 18, 620108 Yekaterinburg, Russia}
\affiliation{Institute of Physics and Technology, Ural Federal University, Mira St. 19, 620002 Yekaterinburg, Russia}

\begin{abstract}
In \linio  Ni$^{3+}$ ion has $d^7$ configuration in cubic crystal field with one electron on double degenerate $e_g$ orbitals, and such ion is considered to be Jahn-Teller (JT) active. However despite the fact, that this compound is an insulator, and hence $d$-electrons are localized, a cooperative JT lattice distortion was not observed. This problem was usually supposed to be resolved by the presence of local JT-distortions that do not order in cooperative JT distorted crystal structure. In the present work DFT+DMFT approach, combining Density Functional Theory with Dynamical Mean-Field Theory, was applied to study electronic and magnetic properties of \linio. In the result, insulating solution with a small energy gap value was obtained in agreement with experimental data. However, in contrast to  previous calculations by other methods,  the symmetry was not broken and the calculated ground state is a thermodynamical mixture of $\alpha d^7 + \beta d^8L $ ($\alpha \approx 60\%, \beta \approx 40\%$) ionic states. The $d^8L $ state is JT inactive and we have found that for the nickel $d^7$ state two configurations with an electron on the Ni $d_{x^2-y^2}$ or $d_{3z^2-r^2}$ orbital have equal statistical weights. So the orbital degeneracy of Ni$^{3+}$ ion is not lifted and that explains the absence of the cooperative JT lattice distortion in this compound. Also, the temperature dependence of inverse magnetic susceptibility of \linio has been calculated  and a good agreement with experimental data was obtained.
\end{abstract}

\maketitle

\section{Introduction}
Starting from the paper of Goodenough~{\em et al.}~\cite{Goodenough1958a}, it is accepted that in \linio the Ni$^{3+}$ ion in $d^7$ configuration contains a single electron on double degenerate $e_g$ orbitals set with a filled $t_{2g}$ orbitals. Correspondingly the Ni$^{3+}$ ion is in the low-spin magnetic state with $S=1/2$.

The ground state of an isolated Ni$^{3+}$ ion is fourfold degenerate: it has twofold orbital and twofold spin degeneracy. 
A standard scenario would be that the orbital degeneracy is resolved by a (cooperative) Jahn-Teller effect, while the spin degeneracy is lifted by magnetic ordering. 
Let us note that, as far as the $e_g$-electrons are concerned, the cooperative Jahn-Teller effect is synonymous with orbital ordering. Thus it can be explained with a purely electronic model, without the consideration of electron-lattice coupling.

\linio does not undergo a Jahn-Teller distortion~\cite{Rougier1995}, and though the measured susceptibility shows some anomalies, it does not seem to develop magnetic long-range order~\cite{Yamaura1996,Hirota1991}.
So this compound presents a problem, where insulator with transition metal ion in Jahn-Teller active configuration remains in the paraorbital and paramagnetic state till the lowest temperatures. 

This problem was supposed to be resolved by the presence of local JT-distortions that do not order in cooperative JT distorted crystal structure.
Local JT distortions have been observed with extended and transmission x-ray-absorption fine structure (EXAFS and XAFS) experiments~\cite{Rougier1995,Nakai1998}.
The absence of the long-range JT distortion around the Ni$^{3+}$ ion was explained with randomly oriented JT orbitals~\cite{Rougier1995,POUILLERIE2001187}; formations of 10~nm sized domains with orbital ordering and local JT distortion within, but with undistorted structure in average~\cite{Chung2005}; random distribution of the Ni$^{2+}$ impurities within \linio~\cite{Petit2006}, charge disproportionation~\cite{Chen2011a} of the Ni$^{3+}$ cations into Ni$^{2+}$ and Ni$^{4+}$.
The problem of the orbital-ordering and the JT effect existence in stoichiometric \linio seems to be still open. 
Magnetic measurements show anomalous magnetic properties of \linio at low temperatures but without long-range magnetic ordering (see for example~\cite{Reynaud2001} and Ref. 2-20 within). Magnetic susceptibility corresponds to a system of $S=1/2$ spins with weak ferromagnetic coupling~\cite{Yamaura1996}.

An electronic structure calculation for \linio within Density Functional Theory results in metallic ground state that contradicts to experimentally observed insulating state with a small energy gap (0.5~eV in~\cite{Molenda2002}, $\approx$~0.4~eV in~\cite{Anisimov1991} and $\geq$~0.1~eV in~\cite{Galakhov1995}). 
This contradiction is an indicator of electronic states localization and importance of electronic Coulomb correlations.
If one takes into account the correlations within DFT+U approach, then insulator solution could be obtained~\cite{Anisimov1991}, but DFT+U method assumes a long-range magnetic and orbital order while experimentally paramagnetic and paraorbital state is observed till the lowest temperatures in \linio. 

DFT+U method corresponds to static mean-field approximation for the Coulomb interaction Hamiltonian ~\cite{Anisimov1991,LDA+U2}. It means that its solution corresponds to the ground state in the form of a single Slater determinant with fixed spin-orbital occupancy values, that breaks symmetry and impose orbital and spin order on the system. Previously published results of calculations within the DFT+U indicates that a low-symmetry JT-distorted structure is the lowest one for \linio~\cite{Chen2011,Chen2012}.
However, one can use Dynamical Mean--Field Theory (DMFT)~\cite{DMFT,DMFT2,DMFT3} approach and obtain the ground state as a thermodynamical mixture of various electronic configurations (Slater determinants). 
Statistical weights of the contributing configurations to the ground state could be computed directly. In the result, one obtains a Green function that describes ground state and excitation spectra for the system under consideration, without breaking symmetry and imposing unnecessary spin and orbital order. Hence it is possible to obtain an insulating solution for one electron in $e_g$ band preserving high symmetry paraorbital and paramagnetic state. This was demonstrated on a model level in~\cite{Poteryaev2008}.

In the present work, we have used {\em ab initio} DFT+DMFT approach, combining Density Functional Theory with Dynamical Mean--Field Theory ~\cite{DFT+DMFT}, to calculate electronic structure, spectral and magnetic properties of  \linio. 
We used two effective non-interacting Hamiltonians in the basis of Wannier functions, constructed as a postprocessing step of DFT calculation of the compound. The first Hamiltonian corresponds to a minimal model in the basis of two Wannier functions, corresponding to the partially filled $e_g$-band. Since it is conventional that Ni$^{3+}$ ion could be not in $d^n$, but in $d^{n+1}L$ configuration, the second Hamiltonian includes in addition oxygen states and takes into account O-$2p$ and partially filled Ni-$e_g$ states hybridization effects in \linio.
The analysis of the calculated two bands Hamiltonian parameters indicates that the triangular lattice models with the nearest neighbors hopping only, used in literature~\cite{Mostovoy2002,Vernay2004a} for the \linio magnetic and orbital orderings description, are oversimplified. The calculated intersite electron transfer energies for the Ni ions are rather long-ranged that was ignored earlier.
In the result of DFT+DMFT calculations for the constructed Hamiltonians, we have obtained small gap insulator in the paramagnetic and paraorbital state in an agreement with experimental data. 
The nickel ions are in the mixed $\alpha d^7 + \beta d^8L $ configuration where the statistical weight of $d^7$ state is 56\% and the weight of $d^8L$ state is 40\% (there is also about 4\% of $d^9L^2$ configuration at 232~K). 
The Ni $d^8L$ state does not assume the appearance JT distortion. Within the $d^7$ state the statistical weight of the configuration with filled Ni $d_{x^2-y^2}$ orbital equals the weight of the configuration with filled Ni $d_{3z^2-r^2}$ orbital. Therefore in the obtained solution there is no prerequisites for the JT distortion arise in \linio.
We have also calculated temperature dependence of magnetic susceptibility that agrees well with experiment.

\section{Methods}

In this work, we have used the  DFT+DMFT calculation procedure described in~\cite{Korotin2008a}. The DFT calculation was done with the Quantum ESPRESSO~\cite{Giannozzi2009} package, PBE exchange-correlation functional, a regular 16x16x16 k-points mesh in the irreducible part of Brillouin zone for reciprocal space integrals, and the energy cutoff values equals 45~Ry and 450~Ry for wavefunctions and charge density respectively. 
The lattice parameters for space group {\em R-3m} were taken a=2.833~\AA\xspace and c=14.215~\AA~\cite{li1993situ}.

\begin{figure}
    \includegraphics[width=\columnwidth]{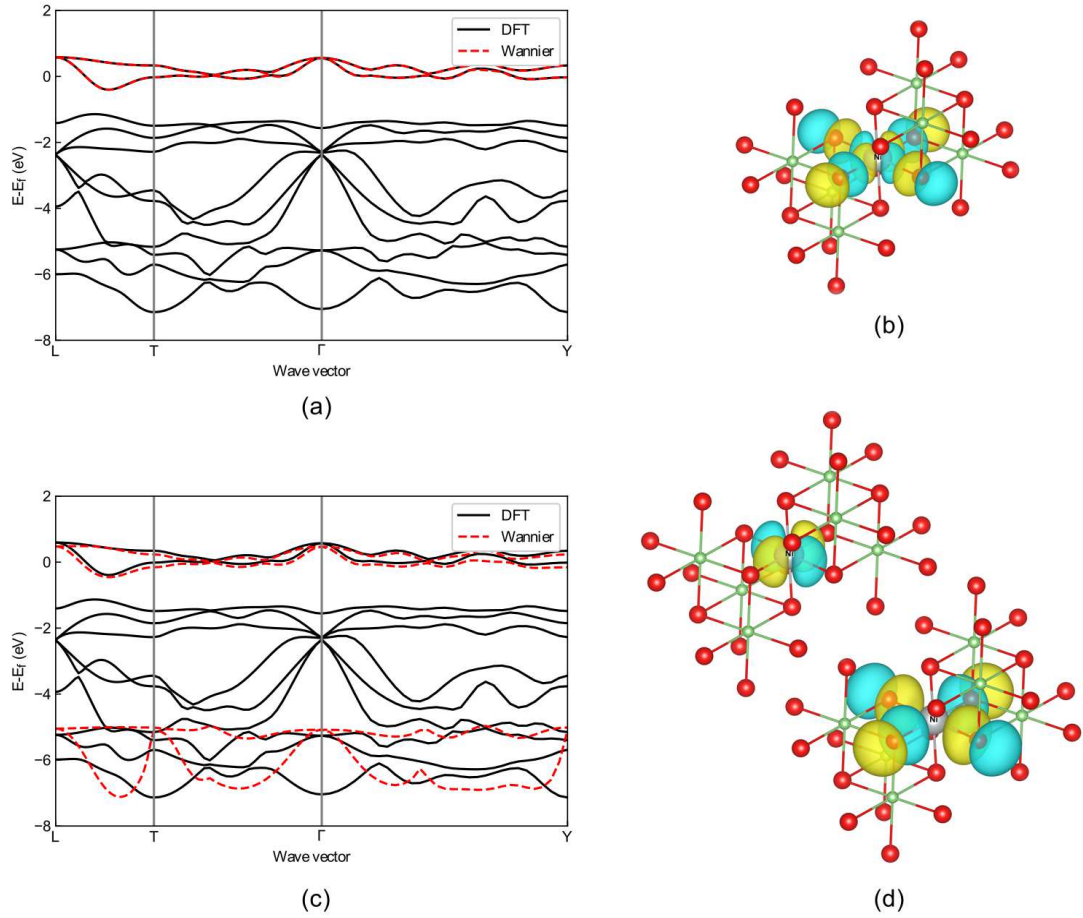}
    \caption{(a) Calculated DFT band structure of \linio (black solid line) and bands obtained from the non-interacting Hamiltonian in the minimal Wannier functions basis (red dashed line). (b) Corresponding WF with the symmetry of Ni-$d_{x^2-y^2}$ orbital. (c) The DFT (black solid line) and the non-interacting Hamiltonian (red dashed line) band structure of \linio for the second basis set, that directly includes oxygen p-states hybridized with Ni-$e_g$. (d) The resulting  Ni-$d_{x^2-y^2}$ and O-$p$ WFs. The two additional basis WFs (Ni-$d_{3z^2-r^2}$ and another O-$p$) are not shown. Red spheres denote oxygen ions, green ones -- Li ions. }
    \label{fig:bands}
\end{figure}

In the crystal field of the ligands octahedron, the Ni-$d$ energy bands are split into filled $t_{2g}$ subband and partially filled $e_g$ subband with the one electron.
In fact, the cubic degeneracy of the Ni $t_{2g}$ states is lifted due to the trigonal crystal structure distortion existing in \linio and the double-degenerate $e_g^{\pi}$ states and non-degenerate $a_{1g}$ states are formed instead of triple degenerate $t_{2g}$ states. 
However, since we consider $t_{2g}$-states as filled, we did not include these states in consideration and only the partially filled $e_g^{\sigma}$ states are taken into account as correlated in DFT+DMFT calculations.
The band structure of \linio calculated within DFT is presented in Fig.~\ref{fig:bands} (a). 
The Fermi level crosses two partially filled $e_g^{\sigma}$ energy bands that are separated from the fully occupied low-energy states formed by O $2p$ and Ni $3d$ $t_{2g}$. 

We used the two different basis sets for the model Hamiltonians to consider the Coulomb correlations in \linio. The first one is a minimal basis of 2 Wannier functions (WF) with the symmetry of Ni $e_g$ orbitals. The WFs were constructed by a projection of Bloch functions with energies in the interval  [-1;1]~eV around the Fermi level on the atomic wavefunctions centered on the Ni ions and having symmetry of Ni $e_g$ orbitals (in details the projection routine is described in~\cite{Korotin2008a}). We did not perform an additional localization procedure here, striving to keep the symmetry of WF unchanged.
The energy bands of the resulting model Hamiltonian and spatial distribution of the basis WFs are shown in Fig.~\ref{fig:bands} (a and b)

Since there is a significant hybridization between nickel and oxygen states in \linio, the two partially filled energy bands, which cross the Fermi level, are formed by a mixture of the Ni-$d$ and O-$p$ orbitals. 
The Wannier functions of the minimal basis (Fig.~\ref{fig:bands} (b)), describing the partially filled energy bands,  are centered on the Ni ion and have a substantial contribution from the $p$-states of the neighboring oxygen ions. Each WF at could be presented as a sum of atomic orbitals $| \phi_{m,n}  \rangle$ $(n=s,p,d,\dots)$ of the neighboring atoms $m$, in the specific case as a sum of Ni-$d$ and the $p$ states of the nearest oxygen ions. Contributions from the other states (Ni-$s$, O-$s$, etc) are negligible.
\begin{equation}
| WF \rangle = \sum_{ n, m } c_{ m ,n } | \phi_{m,n}  \rangle = \sum_{n = d, m = Ni} a_n | \phi_{m,n}  \rangle + \sum_{n = p, m = O} b_n | \phi_{m,n}  \rangle.
\end{equation}

For the minimal basis set describing only 2 energy bands, we estimated each Wannier function's composition ($c_{ m ,n } = \langle \phi_{m,n} | WF \rangle$) as 55\% of Ni-$e_g$ and 45\% of the nearest O-$p$ states.

To consider the charge transfer effect, we build the second model non-interacting Hamiltonian in an extended basis set that includes as Ni-$e_g$ states as well as O-$p$ states hybridized with the former by symmetry. Wishing to keep the number of Wannier functions as small as possible, we took into account only oxygens states that are mostly hybridized with the Ni ones. The Bloch functions with the energies in the full interval [-7;1]~eV were projected on a trial wavefunctions constructed as $| \tilde \phi_d \rangle = \sum_{n = d, m = Ni} a_n | \phi_{m,n}  \rangle $ and $| \tilde \phi_p \rangle = \sum_{n = p, m = O} b_n | \phi_{m,n}  \rangle$, where coefficients $a_n$ and $b_n$ were set the same as for the first basis set Wannier functions. The resulting four WFs (2 WF of Ni-$e_g$ + 2 WFs of O-$p$) and the model non-interacting Hamiltonian band structure are presented in Fig.~\ref{fig:bands} (c and d).

The non-interacting Hamiltonians in the two basis sets were used as input for the DMFT calculation performed within the AMULET package~\cite{AMULET}. Since the Hubbard U parameter depends strongly on WFs spatial distribution and more localized basis assumes larger Coulomb interaction strength~\cite{Anisimov2008}, the U value for the minimal basis (2 WFs of Ni $e_g$-symmetry) was set to $U$ = 4.0~eV~\cite{Laubach2009} and for the second basis set (2 WF of Ni-$e_g$ + 2 WFs of O-$p$) $U$ = 8.0~eV. The Hund exchange parameter $J$ = 0.9~eV was used in both cases. To solve the impurity problem we used continuous-time quantum Monte-Carlo algorithm (CT-QMC)~\cite{gull2011continuous}. 
For the spectral function calculation (Fig.~\ref{fig:spectra}), in the QMC~\cite{QMC} simulations, the inverse temperature value was up to $\beta$=50~eV$^{-1}$ and we used $1.5\times10^6$ Monte Carlo sweeps. 

\begin{figure*}
    \includegraphics[width=\columnwidth]{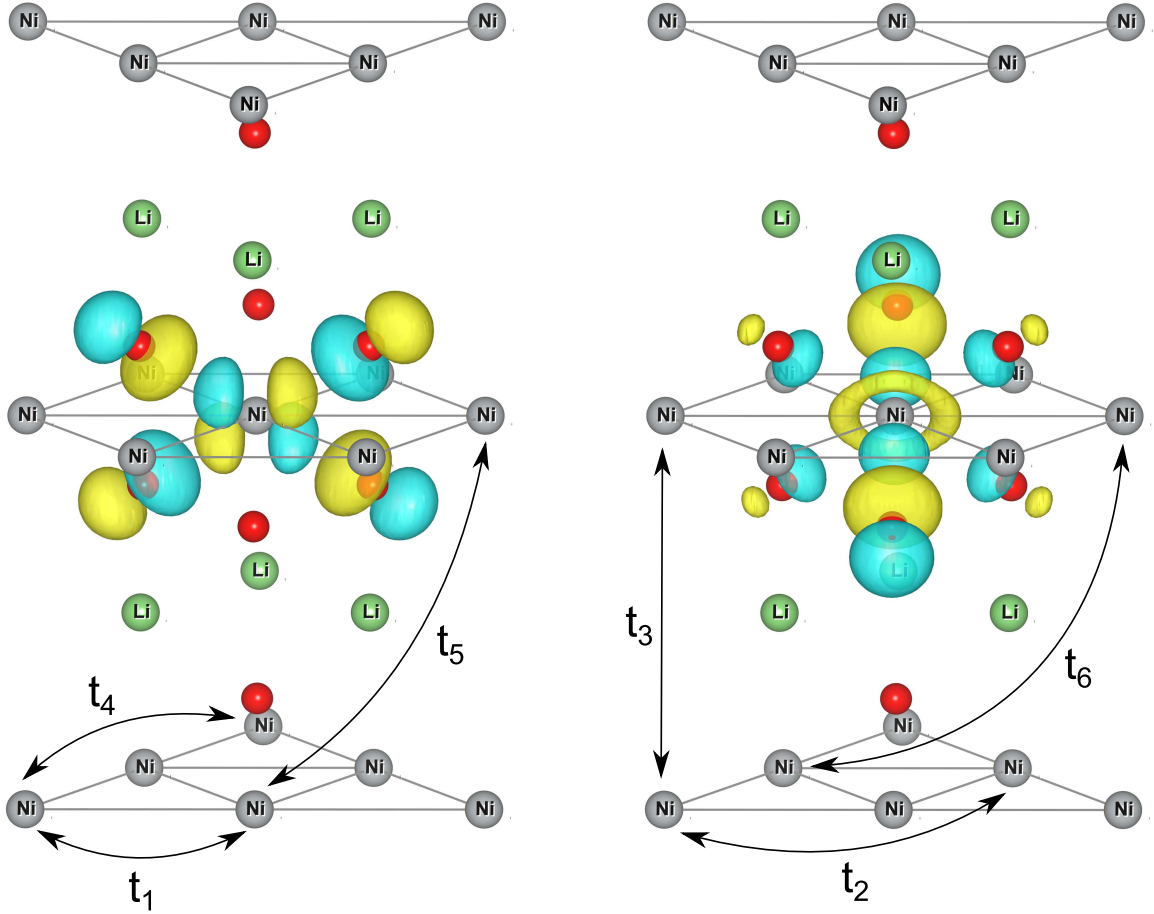}
    \caption{The spatial distribution of  Wannier functions with symmetry of Ni $d_{x^2-y^2}$ orbital (left panel) and $d_{3z^2-r^2}$ orbital (right panel). The Ni ions are denoted with gray balls, the O ions with red balls, and the Li ion with green balls. The directions of the nearest neighboring hoping integrals $t_1 - t_6$ are shown with arrows.}
    \label{fig:wanniers}
\end{figure*}

For the minimal basis set, the calculated kinetic energy of an electron transfer for WF of one Ni site to WF of the neighboring Ni site (i.e., the hopping integral of the effective model Hamiltonian) is long-range, as shown in table~\ref{tab:hopings}. 
The hoping parameters to the 4th nearest Ni neighbor are more significant than the hoppings to the first three nearest neighbors. 
This happens due to an overlap of the neighboring ions WFs on the oxygen sites in between, because of the large contribution to WF from O-$p$ orbitals. 
The significant values for the hoping integrals between the Ni planes ($t_5$ and $t_6$) indicate that if one tries to construct a triangular lattice model for magnetic properties description if \linio as in~\cite{PhysRevLett.81.3527,Mostovoy2002,Hirota1992}, the interlayer Ni-Ni exchange interaction should not be neglected. On the other side, in the extended Ni-$e_g$+O-$p$ basis, the direct Ni-Ni hopings decrease with distance and even for the second nearest neighbor do not exceed 50~eV (see table~\ref{tab:hopings}), however, in this case, the Ni-O electron transfer should be included in any used model.

\begin{table*}
\noindent
\caption{\label{tab:hopings}Calculated kinetic energy (meV) of an electron transfer $t_n^{ij}$ for the $i$-th $e_g$-like WF of one Ni site to the $j$-th $e_g$-like WF of the neighboring Ni site. The electron hoping directions $t_n$ are shown in Fig.~\ref{fig:wanniers}.}
\begin{tabular}{|c|c|c|c|c|c|c|}
\hline
Basis & $t_1^{i,j}$ & $t_2^{i,j}$ & $t_3^{i,j}$ & $t_4^{i,j}$ & $t_5^{i,j}$ & $t_6^{i,j}$ \\
\hline
Ni-$e_g$ &
$\left( \begin{array}{cc}
47 & 0 \\
0 & -16
\end{array} \right) $& 
$\left( \begin{array}{cc}
 10 & -6 \\
 -6 & 4
\end{array} \right) $&
$\left( \begin{array}{cc}
-1 & 0 \\
 0 & 9
\end{array} \right) $&
$\left( \begin{array}{cc}
17 & -42 \\
-42 & 66
\end{array} \right) $&
$\left( \begin{array}{cc}
49 & 0 \\
0 & -19
\end{array} \right) $&
$\left( \begin{array}{cc}
-23 &-13 \\
-13 & -10
\end{array} \right) $
\\

\hline
Ni-$e_g$+O-$p$ &
$\left( \begin{array}{cc}
42 & 0 \\
0 & -66
\end{array} \right) $& 
$\left( \begin{array}{cc}
 17 & -16 \\
 -16 & -1
\end{array} \right) $&
$\left( \begin{array}{cc}
20 & 0 \\
 0 & -15
\end{array} \right) $&
$\left( \begin{array}{cc}
12 & 10 \\
10 & 0
\end{array} \right) $&
$\left( \begin{array}{cc}
-16 & 0 \\
0 & -10
\end{array} \right) $&
$\left( \begin{array}{cc}
-3 & -6 \\
-6 & 14
\end{array} \right) $
\\

\hline

\end{tabular}

\end{table*}

\section{Results and Discussion}

\begin{figure}
    \includegraphics[width=\columnwidth]{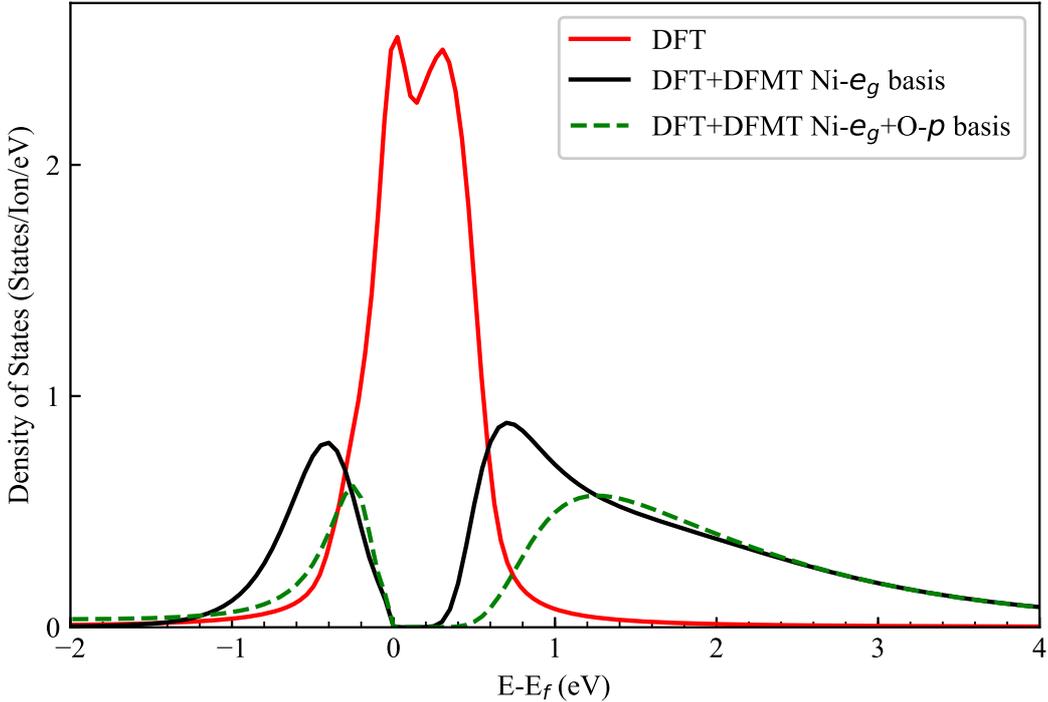}
    \caption{Calculated density of states for $e_g$ electrons of \linio. Red solid  line -- DOS obtained within DFT calculation, black solid line –- DOS calculated within DFT+DMFT approach in the minimal basis of 2 WFs, green dashed line –- DOS calculated within DFT+DMFT approach in the basis of Ni-$e_eg$ + O-$p$ WFs.}
    \label{fig:spectra}
\end{figure}

The density of states of the nickel $e_g$ electrons in \linio calculated within the DFT+DMFT approach for the temperature of 232~K are presented in Fig.~\ref{fig:spectra}. 
We have obtained an insulating solution with the energy gap value from 0.3~eV (the minimal basis) to 0.45~eV (the second basis set) that is in agreement with experimental values 0.4-0.5~eV~\cite{Anisimov1991,Molenda2002}. 
The used method does not enforce any long-range magnetic or orbital ordering, and the obtained solution is paramagnetic and paraorbital. The obtained mean value for the total magnetization operator $\langle s_z \rangle$ is zero in the whole used temperature range 232..1160~K, therefore there is no evidence of a magnetic ordering.

The hybridization expansion CT-QMC solver provides the site-reduced statistical operator (density matrix)~\cite{werner2007high}. This quantity describes the probability of finding an atom in a particular many-body state and an expectation value of any local operator can be easily obtained from it. Therefore, this instrument is well suited to analyze a statistical probability of the various atomic configurations of the Ni ion.

In the calculated ground state, the two Ni $d_{x^2-y^2}$  and $d_{3z^2-r^2}$ orbitals are equally filled and the corresponding configurations have the same statistical weights.
In the minimal basis of 2 WFs each Wannier function is filled with 0.5 electrons in a mean or is totally filled with the probability of 50\%. The second, extended basis of Ni-$e_g$ and O-$p$ WFs, gives a more complex result. Due to the charge-transfer effect taken into account, the configuration $d^8L$ with the totally filled $e_g$ subshell has the statistical weight equal 40\%. This electronic configuration is Jahn-Teller inactive. The $d^7$ configuration has the weight of 56\% and $d^9L^2$ takes the rest 4\%. Even for the second basis, the to WFs of Ni-$e_g$ symmetry have equals occupation number and the same statistical weight.
Consequently, on an average, both $e_g$ orbitals are degenerate and are filled equivalently, therefore, there are no preconditions for the appearance of the Jahn-Teller lattice distortion.

\begin{figure}
\includegraphics[width=\columnwidth]{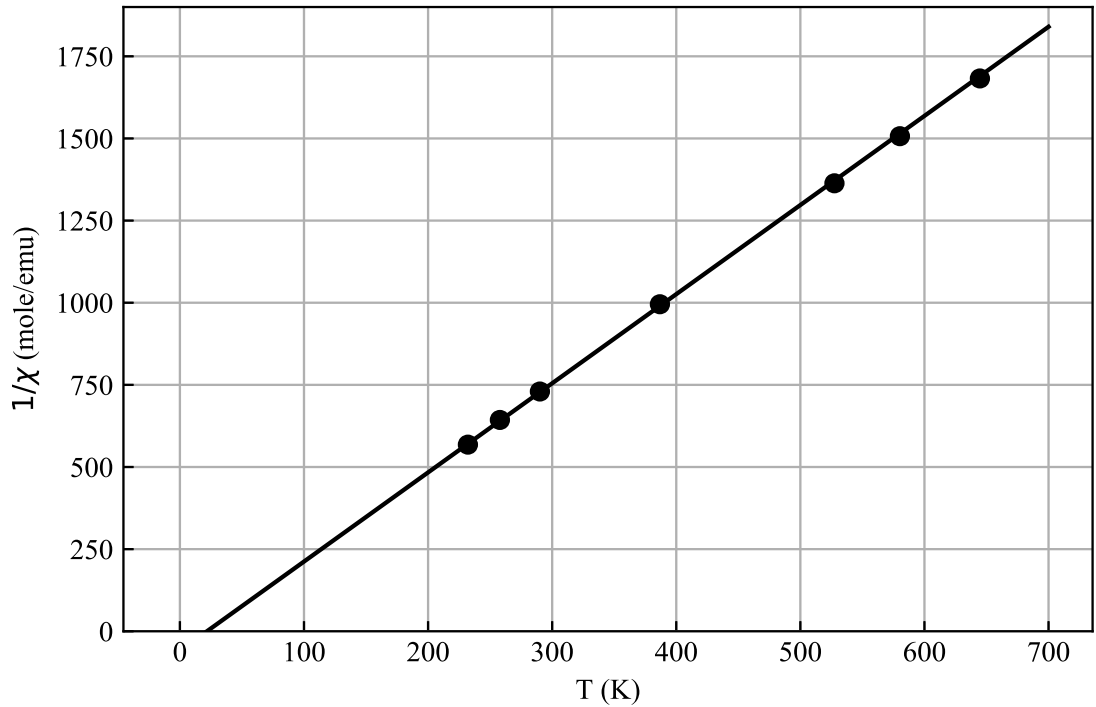}
\caption{Calculated inverse magnetic susceptibility for \linio. The black dots -- calculated values, the black solid line is plotted using the least squares approximation method.}
\label{fig:inv_chi}
\end{figure}

To study magnetic properties of \linio, the temperature dependence of magnetic susceptibility has been calculated in DFT+DMFT method. It was done by applying small magnetic field on Ni ions and calculating resulting spin polarization. The ratio of the polarisation value
to the magnetic field value is susceptibility. Result is shown in Fig.~\ref{fig:inv_chi}.
From the $\chi^{-1}(T)$ dependence the effective magnetic moment of Ni ion and the Curie-Weiss parameter $\theta$ were calculated. 
Calculated value of $\mu_{eff} = 1.22~\mu_B$ is slightly underestimated comparing with with experimental values for the effective moment (1.91~$\mu_B$~\cite{Yamaura1996}, 2.1~$\mu_B$~\cite{Sugiyama2010,Bonda2008,Reimers1993},) but both corresponds to formal spin state $S=1/2$.
The calculated Curie-Weiss parameter $\theta = 22~K$ is in a good agreement with experimental values $\theta$ (19~K~\cite{Reynaud2000}, 26~K~\cite{Reimers1993}, 29~K~\cite{Bonda2008}, 41~K~\cite{Sugiyama2010}). The positive sign of calculated $\theta$ confirms the weak ferromagnetic coupling between $S=1/2$ spins in \linio.

\section{Conclusion}
In DFT+DMFT calculations we have obtained the paramagnetic paraorbital insulating ground state for the \linio. The obtained ground state is the mixture of 56\% $d^7$, 40\% $d^8L$ and 4\% of $d^9L^2$ configurations of the Ni ion. 
The two $e_g$ orbitals have the same average occupancy in all these configurations, therefore even within the $d^7$ configuration a prerequisite of Jahn-Teller distortion of the ligands octahedron is absent.
Within the same approach, the magnetic susceptibility dependence on temperature has been computed. 
Calculated Curie-Weiss parameters $\mu_{eff}$ and $\theta$ are in agreement with available experimental data.

\section{Acknowledgments}
This work was supported by Russian Science Foundation (Project 14-22-00004). D.K. is grateful to A.I.~Poteryaev for valuable discussions.

\bibliography{main}

\end{document}